%% file: 02main.tex
\documentclass[10pt,a4paper,twoside]{article}

\usepackage[dvips]{epsfig}

\usepackage{setspace}

\usepackage{supertabular}
\usepackage{lscape}

\usepackage{fancyhdr}
\newcommand {\sect}[1]{\section{#1} \rule[5mm]{1\headwidth}{0.5mm}}

\makeatletter
\@addtoreset{figure}{section}

\@addtoreset{equation}{section}

\@addtoreset{table}{section}

\makeatother

\def\gappeq{\mathrel{ \rlap{\raise.5ex\hbox{$>$}}
                      {\lower.5ex\hbox{$\sim$}}  } }
\def\lappeq{\mathrel{ \rlap{\raise.5ex\hbox{$<$}}
                      {\lower.5ex\hbox{$\sim$}}  } }

\begin{document}

\pagenumbering{roman}

\thispagestyle{empty}

\begin{flushright}
CERN-AB-2006-004 \\
DESY 06-001 \\
ILC-Asia-2005-26\\
JAI-2006-001\\
KEK Report 2005-9\\
SLAC-R-796 \\
UT-ICEPP 05-04 \\
~ ~
\end{flushright}
\vspace{0.8cm}

\begin{center}
{\bf \Huge ATF2 Proposal}\\
\vspace{0.4cm}
{\bf \Large Vol. 2}
\vspace{1cm}

{\bf \Large ATF2 Group}

February 13, 2006\\
\end{center}

\input{preface.tex}

\cleardoublepage
\sect{Introduction}\label{sec:intro}
\input{introduction.tex}

\cleardoublepage
\sect{Timelines}\label{sec:timeline}
\input{timeline.tex}

\cleardoublepage
\sect{Organization}\label{sec:organization}
\input{organization.tex}

\cleardoublepage
\sect{Cost Estimation}\label{sec:cost}
\input{cost.tex}

\cleardoublepage
\sect{International Contributions}\label{sec:contribution}
\input{contribution.tex}

\cleardoublepage
\input{postscript.tex}

\end{document}

%% file: preface.tex
\clearpage
\thispagestyle{empty}
~ ~


~ ~
\cleardoublepage
\input{preface/author.tex}

\cleardoublepage
\tableofcontents
\cleardoublepage
\listoffigures
\listoftables

\pagenumbering{arabic}
\setcounter{page}{1}

%% file: preface/author.tex
\begin{center}

 Boris Ivanovich Grishanov,
 Pavel Logachev,
 Fedor Podgorny,
 Valery Telnov\\
 {\it (BINP SB RAS, Novosibirsk)} 

 Deepa Angal-Kalinin,
 James Jones,
 Alexander Kalinin\\
 {\it (CCLRC/DL/ASTeC,Daresbury, Warrington, Cheshire)} 

 Olivier Napoly,
 Jacques Payet\\
 {\it (CEA/DSM/DAPNIA, Gif-sur-Yvette)} 

 Hans-Heinrich Braun,
 Daniel Schulte,
 Frank Zimmermann\\
 {\it (CERN, Geneva)} 

 Robert Appleby,
 Roger Barlow, 
 Ian Bailey, 
 Leo Jenner, 
 Roger Jones, 
 German Kourevlev\\
 {\it (The Cockcroft Institute, Daresbury, Warrington, Cheshire)} 

 Eckhard Elsen,
 Vladimir Vogel,
 Nick Walker\\
 {\it (DESY, Hamburg)} 
 
 Nikolay Solyak,
 Manfred Wendt\\
 {\it (Fermilab, Batavia, Illinois)} 

 Tohru Takahashi\\
 {\it (Hiroshima University, Higashi-Hiroshima)} 

 Jie~Gao, 
 Weibin~Liu, 
 Guo-Xi~Pei, 
 Jiu-Qing~Wang\\
 {\it (IHEP, Beijing)} 

 Nicolas Delerue,
 Sudhir Dixit,
 David Howell,
 Armin Reichold,
 David Urner\\
 {\it (John Adams Institute at Oxford University)} 

 Alessio Bosco,
 Ilya~Agapov,
 Grahame~A.~Blair\footnotemark[1],
 Gary Boorman,
 John Carter,
 Chafik Driouichi,
 Michael Price\\
 {\it (John Adams Institute at Royal Holloway, Univ. of London)} 

 Sakae~Araki,
 Hitoshi~Hayano,
 Yasuo~Higashi,
 Yosuke~Honda,
 Ken-ichi~Kanazawa,
 Kiyoshi~Kubo,
 Tatsuya~Kume,
 Masao~Kuriki,
 Shigeru~Kuroda,
 Mika~Masuzawa,
 Takashi~Naito,\\
 Toshiyuki Okugi,
 Ryuhei Sugahara,
 Toshiaki Tauchi,\footnotemark[1]
 Nobuhiro~Terunuma,\\
 Nobu~Toge,
 Junji Urakawa,
 Hiroshi Yamaoka,
 Kaoru Yokoya\\
 {\it (KEK, Ibaraki)} 

 Yoshihisa Iwashita,
 Takanori Mihara\\
 {\it (Kyoto ICR, Uji, Kyoto)} 

 \renewcommand{\thefootnote}{\fnsymbol{footnote}}
 Maria Alabau Pons\footnotemark[1]\footnotetext[1]{LAL, Orsay and IFIC, Valencia},
 Philip Bambade, 
 Olivier Dadoun\\
 {\it (LAL, Orsay)} 
 \renewcommand{\thefootnote}{\arabic{footnote}}

 \clearpage

 Beno\^it Bolzon, 
 Nicolas Geffroy, 
 Andrea Jeremie, 
 Yannis Karyotakis\\
 {\it (LAPP, Annecy)} 

 Andy Wolski\\
 {\it (LBL, Berkeley, California)} 

 Jeff Gronberg\\
 {\it (LLNL, Livermore, California)} 

 Stewart Takashi Boogert,
 Alexey Liapine,
 Stephen Malton,
 David J. Miller,
 Matthew Wing\\
 {\it (University College London, London)} 

 Masayuki Kumada\\
 {\it (NIRS, Chiba-shi)} 

 Samuel Danagoulian,
 Sekazi Mtingwa\\
 {\it (North Carolina A\&T State University, North Carolina)} 

 Eric Torrence\\
 {\it (University of Oregon, Eugene, Oregon)} 

 Jinhyuk Choi,
 Jung-Yun Huang,
 Heung Sik Kang,
 Eun-San Kim,
 Seunghwan Kim,
 In Soo Ko\\
 {\it (Pohang Accelerator Laboratory)} 

 Philip Burrows,
 Glenn Christian,
 Christine Clarke,
 Anthony Hartin,
 Hamid Dabiri Khah,\\
 Stephen Molloy,
 Glen~White\\
 {\it (Queen Mary University of London, London)} 

 Karl Bane,
 Axel Brachmann,
 Thomas Himel,
 Thomas Markiewicz,
 Janice Nelson,
 Yuri Nosochkov,\\
 Nan Phinney,
 Mauro Torino Francesco Pivi,
 Tor Raubenheimer,
 Marc Ross,
 Robert Ruland,\\
 Andrei Seryi\footnotemark[1],
 Cherrill M. Spencer,
 Peter Tenenbaum,
 Mark Woodley\\
 {\it (SLAC, Menlo Park, California)} 

 Sachio Komamiya, 
 Tomoyuki Sanuki\footnotemark[1]\footnotetext[1]{The Editorial Board},
 Taikan Suehara\\
 {\it (University of Tokyo, Tokyo)} 
 
\end{center}

%% file: introduction.tex
\input{introduction/introduction.tex}

%% file: introduction/introduction.tex
For achieving the high luminosity required at the International Linear 
Collider (ILC), it is critical to focus the beams to nanometer size with 
the ILC Beam Delivery System (BDS), and to maintain the beam collision with 
a nanometer-scale stability.

To establish the technologies associated with this ultra-high precision beam 
handling, it has been proposed to implement an ILC-like final focus optics 
in an extension of the existing extraction beamline of ATF at KEK. 
The ATF is considered to be the best platform for this exercise, since it 
provides an adequate ultra-low emittance electron beam in a manner 
dedicated to the development of ILC.

The two major goals for this facility, called ATF2, are : (A) Achievement of 
a 37~nm beam size, and (B) control of beam position down to 2~nm level.
The scientific justification for the ATF2 project and its technical design
have been described in Volume 1 of the ATF2 Proposal \cite{atf2vol1}. 
We present here Volume 2 of the ATF2 Proposal, in which we present
specifics of the construction plans and the group organization to execute
the research programs at ATF2.  

The sections in this report have been authored 
by relevant ATF2 subgroups within the International ATF Collaboration. 
The time line of the project is described in Section \ref{sec:timeline}.
Section \ref{sec:organization} discuss the structure of the international collaboration.
Sections \ref{sec:cost} and \ref{sec:contribution} discuss 
budget considerations, which are presented
as well as
the design and construction tasks 
to be shared by the international collaboration at ATF2. 
Concluding remarks 
have been contributed 
by Dr. Ewan Paterson, 
Chair of the International Collaboration Board of the ATF collaboration.

%% file: timeline.tex
\input{timeline/yokoya.tex}

%% file: timeline/yokoya.tex
The purpose of ATF2 is to address two major challenges of the ILC BDS:
focusing the beams to nanometer size and providing sub-nanometer stability. 
More specifically, as described in the Volume 1\cite{atf2vol1}, the goals of the ATF2 are:
\begin{itemize}
\item[(A)] Achievement of a 37~nm beam size
 \begin{itemize}
  \item[(A1)] Demonstration of a compact final focus system based on a local chromaticity correction scheme
  \item[(A2)] Maintenance of the small beam size 
 \end{itemize}
\item[(B)] Control of the beam position
  \begin{itemize}
   \item[(B1)] Demonstration of beam orbit stabilization with nano-meter precision at IP.
   \item[(B2)] Establishment of a beam jitter controlling technique at the nano-meter level with an ILC-like beam 
  \end{itemize}
\end{itemize}

In defining the timeline of the ATF2, severals facts have to be taken into consideration.

~~

The official plan proposed by the ILC-GDE (Global Design Effort) is to complete the RDR (Reference Design Report) 
by the end of 2006 and to write the TDR (Technical Design Report) after two or more years.
The scientific goals of the ATF2 have little to do with the RDR. 
For example, ATF2 will not exert a large influence on the choice of the crossing angle.
On the other hand ATF2 can have a big impact on the TDR. The commissioning process of the ILC Final Focus System
is very much complicated, as expected from the experience of the SLC. We can learn from the ATF2 
what sort of diagnostics system and tuning algorithm are useful for minimizing the process. 
Thus, ATF2 should be completed well before the TDR, 
though we do not know exactly when TDR will be finished. 



The timelines 
take into consideration
the international nature of the design discussion, fabrication
and procurement processes of the hardware components and their testing. 

\begin{figure}[tb]
\begin{center}
\includegraphics[width=\textwidth]{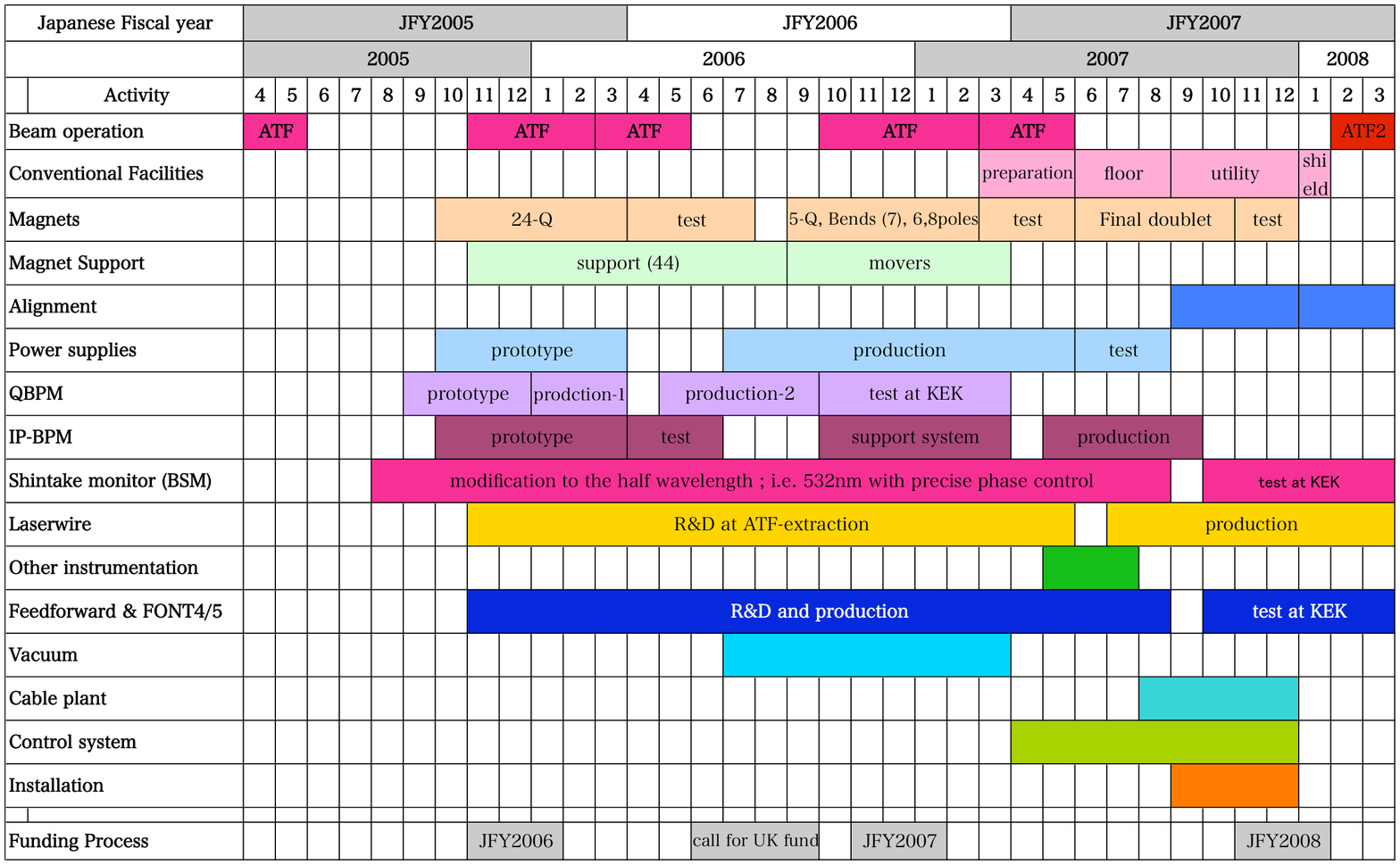}
\caption{Outline of the ATF2 time schedule.}
\label{fig:ATF2schedule}
\end{center}
\end{figure}

The schedule of the ATF must also be taken into account. As described in Volume 1 of the 
present proposal \cite{atf2vol1}, 
ATF2 requires a refurbishment of the floor of the ATF building, 
which will take a few months.
In order not to reduce the precious machine time of the ATF, the refurbishment work 
must be done during a summer shutdown. Also, the experimental studies at ATF2 after its 
construction must be carefully scheduled in a consistent way with other ATF studies.
One must keep in mind that there are still many subjects to be studied at the ATF to be done
in parallel with the ATF2, such as 
the development of diagnostics devices, beam-dynamics studies, etc, 

We decided to adopt the following schedule, while taking the above facts into consideration:
\begin{itemize}
\item  All of the major components, such as magnets, power supplies, BPMs and vacuum components,
	must be completed by June, 2007.
\item  The floor refurbishment and subsequent construction of the shield are to be done
    during the summer shutdown in 2007 (mid-June to October).
\item  Installation of all the components needed for the ATF2 Goal {\bf A} will be done 
	in September through December, 2007.
\item  The first beam operation of ATF2 is expected in February, 2008.
	Then, at least ATF2 Goal {\bf A} (achievement of 37~nm beam size) 
	and presumably Goal {\bf B1} (orbit stabilization to nanometer level) can be achieved 
	by the time of TDR completion.
\end{itemize}
Fig.~\ref{fig:ATF2schedule} shows an outline of the schedule.


Our present schedule leads us to completion of the ATF2 installation by early 2008. 
This will be followed by commissioning of the beamline,
and subsequent efforts towards achieving Goal {\bf A} as rapidly as possible. 
We realize that it is highly desirable to accumulate a sufficient amount of real-life experiences 
from this beamline so as to provide the Technical Design Report of ILC 
with substantially practical inputs from an operational standpoint. 
It is also anticipated that even during the construction of ILC, 
continued beam operation of ATF2 is expected 
to offer opportunities of further developing beam tuning
as well as a training ground for a younger generation of accelerator physicists and engineers.

Moreover, the experience with tuning procedures for ATF2 will also
serve to minimize the commissioning period of the ILC FFS. Thus, the study at ATF2 will
continue even after the start of ILC construction. 
Even during the machine operation of ILC,
ATF2 may serve as a test bed for solving possible problems at ILC and for new ideas. 
It is also within the scope of ATF2 to test a laser system for the gamma-gamma collider.
Thus, ATF2 is a long-term project.

The details of the schedule are shown in Fig.~\ref{fig:ATF2schedule1}.
\begin{figure}[htb]
\begin{center}
\rotatebox{90}{
\includegraphics[width=20cm]{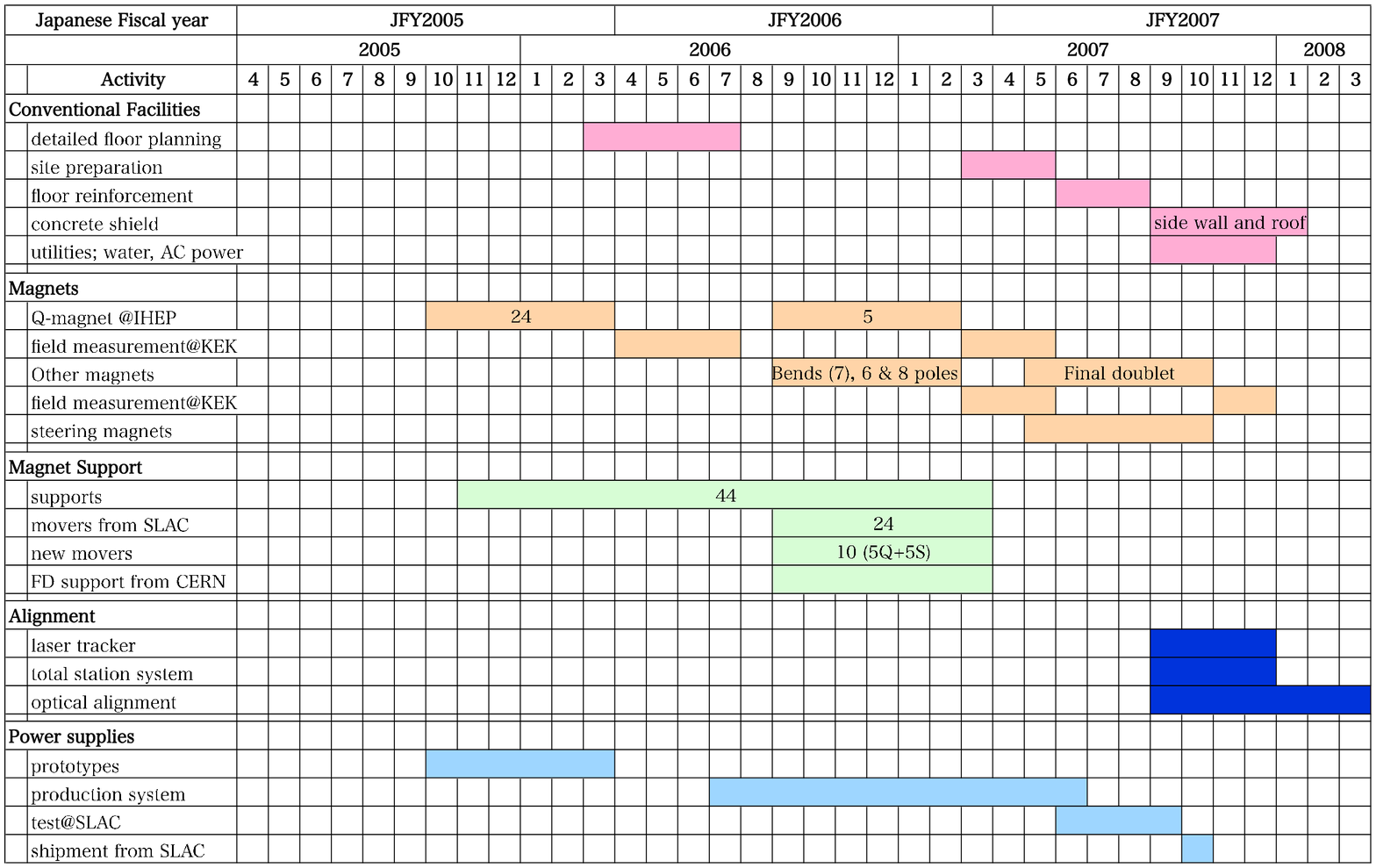}}
\caption{Time schedule of ATF2.}
\label{fig:ATF2schedule1}
\end{center}
\end{figure}
\addtocounter{figure}{-1}\begin{figure}[htb]
\begin{center}
\rotatebox{90}{
\includegraphics[width=20cm]{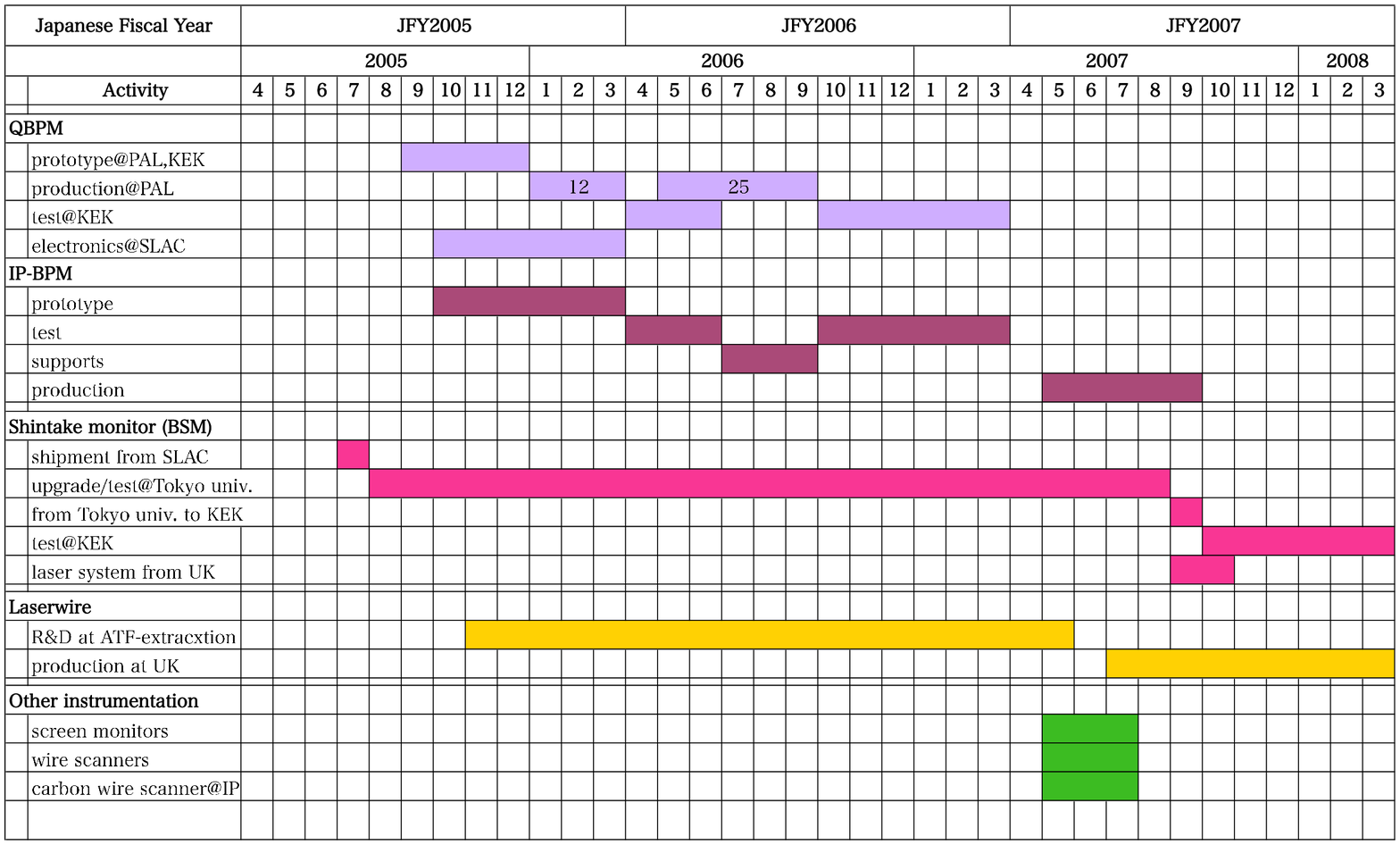}}
\caption{({\it continued})
}
\label{fig:ATF2schedule2}
\end{center}
\end{figure}
\addtocounter{figure}{-1}
\begin{figure}[htb]
\begin{center}
\rotatebox{90}{
\includegraphics[width=20cm]{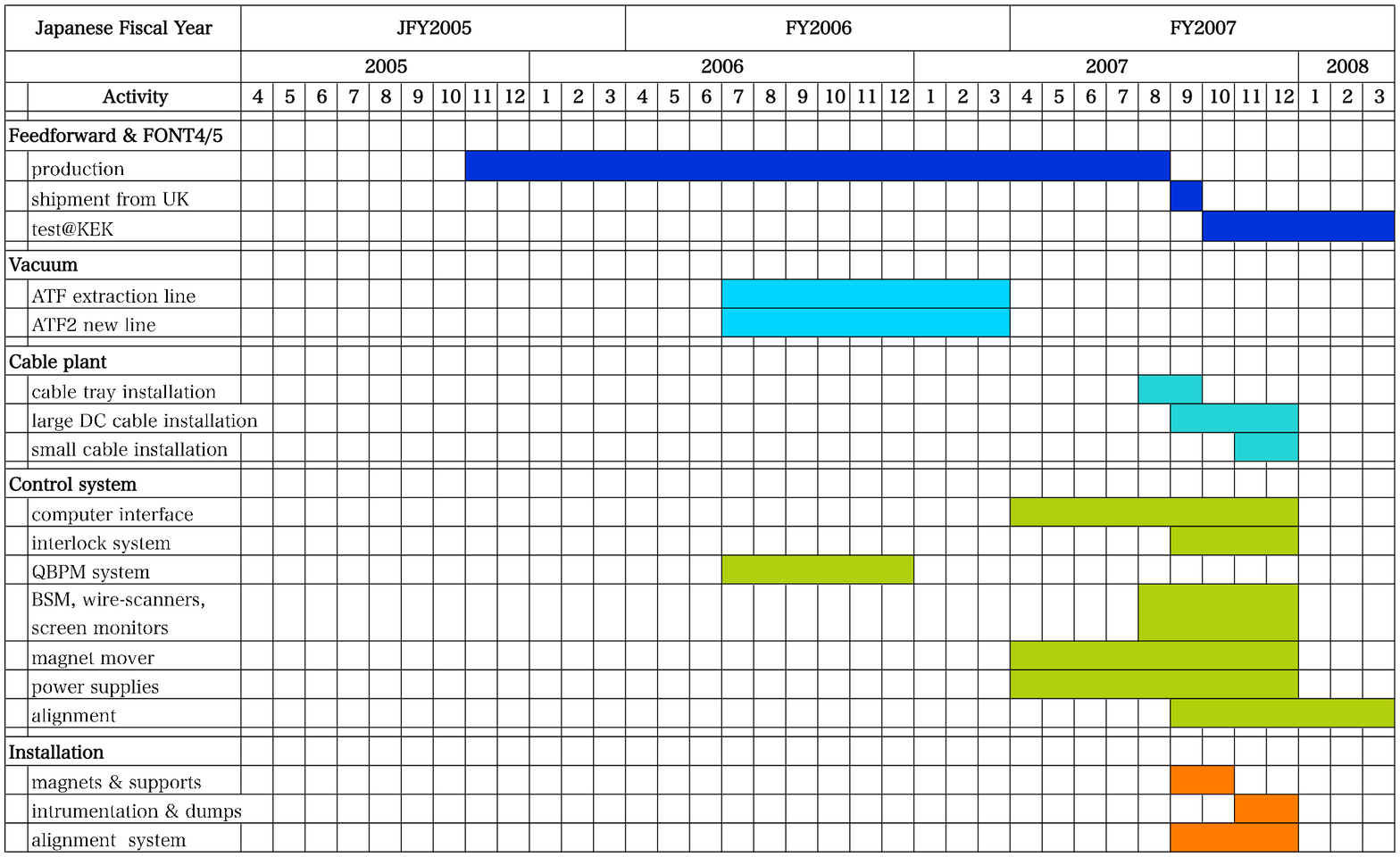}}
\caption{({\it continued})
}
\label{fig:ATF2schedule3}
\end{center}
\end{figure}

%% file: organization.tex
\input{organization/urakawa.tex}

%% file: organization/urakawa.tex
\subsection{Introduction}
We have established an International Collaboration of ATF with many institutes;
it was launched on August 1st, 2005. 
Now, several additional institutes are preparing to join this collaboration. 

The International Collaboration of ATF is based on the Memorandum of Understanding (MoU),
which defines the organization of the international collaboration to carry out research programs 
at ATF accelerator complex, so as to maximally contribute to the world design and development 
efforts in the areas of particle sources, damping rings, beam focusing and beam instrumentation 
towards the International Linear Collider (ILC) project. 
We are adding MoU, a chart of the organization and related material 
onto new ATF Homepage (http://atf.kek.jp/).

\subsection{ATF and ATF2}
The design and preparations of the ATF2 project have been proceeded for construction by several 
international groups (design, magnet, instrumentation, vacuum, alignment and final focus). 
ATF2 group has been organized under the ATF international collaboration since 2005. 
The member list of the organization will be shown in (http://atf.kek.jp/) with the approval of ICB. 
As is evident  from the ATF MoU, 
the construction and operation of ATF2 will be executed 
under the leadership of the Deputy and Sub-Deputy for the ATF2 project 
within the framework of the ATF International Collaboration. 
The management of activities of ATF2 will be carried out under the supervising bodies, 
as described in the subsequent sub-sections.

~~

In the following
we describe a simplified explanation of the ATF MoU and 
additional decisions after the 1st International Collaboration Board (ICB) meeting.

\subsection{Organization of ATF}
To execute the scientific programs at ATF, the following bodies were instituted:
\begin{itemize}
\item International Collaboration Board (ICB)
\item Technical Board (TB)
\item Spokesperson (SP) with his/her Deputies
\item System/Group Coordinators (SGCs)
\end{itemize}

\subsubsection{International Collaboration Board (ICB)}
The International Collaboration Board (ICB) is a decision-making body for executive matters 
related to the ATF collaboration. Each collaborating institute can delegate one member to the 
ICB.   In addition, the ICB is joined by the three GDE Regional Directors, who represent Asian, 
North American and European regions.

One of the members of the ICB is to serve as the ICB Chair. The nomination of the ICB Chair is 
done through mutual voting by all of the ICB members.

According to the 1st ICB meeting at Snowmass 2005, an ICB meeting will be held approximately 
once per year. Research programs and near-term schedules of the machine operation will be  
posted to the ATF homepage, usually for the upcoming two weeks. If some collaborators  have 
a need to change the schedule, SP or Deputies have to coordinate it.

\subsubsection{Technical Board (TB)}
The TB consists of approximately 4 $\sim$ 5 members 
from each of the Asian, North American and European regions. 
The members of the ATF Technical Board (TB) are nominated and appointed by the ICB. 
The TB conducts the following tasks:
\begin{itemize}
\item At the request of ICB, assist the Spokesperson in formulating the ATF Annual Activity Plan, 
which outlines the activity plans of ATF, including the budget and beam-time allocation for 
each Japanese fiscal year.
\item Assist the ICB in assessing the scientific progress that is being made by the ATF 
collaboration.
\end{itemize}
The SP  has three Deputies and serves as the TB Chair for efficient management. 
Usually, the schedule of the ATF machine operation is divided 
to two blocks of  operating time per year with about four months of a long summer shutdown. 
One block is from the middle of October to December, 
and the other is from the middle of January to the middle of June. 
Since there are many collaborators, 
we propose two meetings of TB per year, in December and May, 
to review and recommend the research programs at ATF. 

\subsubsection{Spokesperson (SP)}
The SP carries out the following tasks:
\begin{itemize}
\item Direct and coordinate the work required at ATF in accordance with the ATF Annual 
Activity Plan.
\item Report on progress made by the collaboration to the ICB and the director of KEK.
\item Report on matters related to the KEK budget and KEK properties to the director of KEK.
\end{itemize}

To carry out these tasks, the SP will:
\begin{itemize}
\item Appoint, with the approval of ICB, up to three Deputies to assist in his/her tasks in the 
areas of 
 \begin{itemize}
 \item Beam operation,
 \item Hardware maintenance, and
 \item Design, construction and commissioning of ATF2.
 \end{itemize}
\item Appoint, with the approval of ICB, the System/Group Coordinators (SGCs) on critical 
ATF/ATF2 subsystems and study programs.
\item Organize a "Coordination Group" with the Deputies and System/Group Coordinators for 
coordinating the details of the operation and development at ATF on a daily (during the 
beam operation period) or weekly (during the maintenance and construction period) basis.
\end{itemize}
The report and the discussion are usually carried out through the ATF Homepage (http://atf.kek.jp/) 
with members of ICB, TB and SGCs.

\subsubsection{System/Group Coordinators (SGCs)}
\begin{itemize}
\item The System/Group Coordinators were appointed by the Spokesperson with the approval of ICB. 
\item The appointment of the System/Group Coordinators was made in a manner consistent with the ATF Annual Activity Plan. 
\item The System/Group Coordinators coordinate the tasks charged to the assigned Systems or Groups, and assist the Spokespersons and the Deputies coordinate the ATF research programs.
\end{itemize}
In the case of small study groups with less than 5 members, 
the Spokesperson or the Deputies may 
assume the role of its Coordinator on an acting basis.

\begin{figure}[thb]
\begin{center}
\includegraphics[width=14cm]{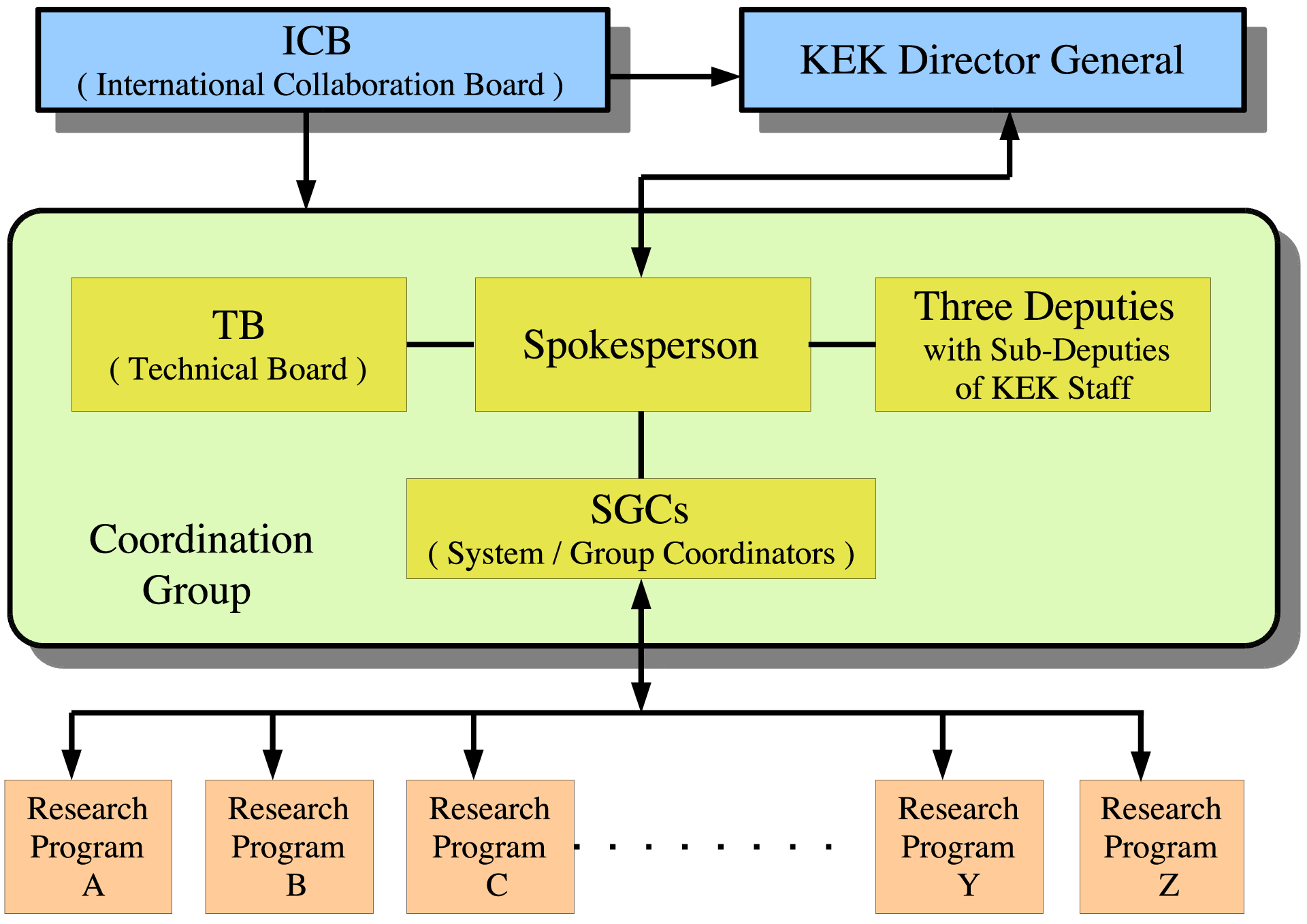}
\caption{Simplified chart of ATF organization.}
\label{fig:organization}
\end{center}
\end{figure}

A simplified organization chart is shown in Fig.~\ref{fig:organization}.

\subsection{Execution of Research Programs at ATF}
The ATF Spokesperson (SP) supervises the construction and operation programs of ATF in 
accordance with the ATF Annual Activity Plan, as approved by the ICB. The ATF SP can delegate 
part of his/her tasks to the Deputies and System/Group Coordinators, as deemed appropriate. The 
planning and execution of the construction and operation programs of ATF must be conducted 
in a manner consistent with the Japanese laws, KEK internal regulations and other rules 
applicable in the cases of activities by non-KEK members of the collaboration. Details of the group 
structures are to be formulated in ways optimized in accordance with the technical nature of 
each program in question, e.g.. activities such as design, simulation, testing, construction, 
commissioning, operation, and investigation.

Additional agreements concerning matters related to 
execution of non-KEK budget, management of non-KEK properties at the premise of ATF, 
together with the handling of KEK properties at the premises outside KEK, 
are to be individually dealt with in Annexes attached to the ATF MoU.

The detailed coordination of the beam-time allocation will be done by the Coordination Group.

An institute that has not signed onto the ATF MoU may participate in 
any part of the research programs at ATF, 
after authorization of the ATF SP, as long as the proposed activity is:
\begin{itemize}
\item within the ATF Annual Activity Plan, 
\item under the consent by the relevant the Deputies or System/Group Coordinators, and 
\item promptly reported to the ICB.
\end{itemize}
Should a conflict of interests occur among the members or Work Groups within the collaboration, 
the Spokesperson will make the best efforts to resolve it in an amicable manner. When a suitable 
resolution cannot be reached, the Spokesperson will bring the matter to the ICB for further 
negotiation towards a resolution.

\subsection{Membership of the Collaboration}
Membership of new institutes for the ATF collaboration is subject to approval of the ICB. 
Institutes that desire to join the ATF collaboration shall submit a proposal to the Spokesperson, 
who will relay the matter to ICB.

The withdrawal of a member institute from the ATF collaboration is subject 
to acknowledgment by the ICB. 
Institutes that desire to withdraw from the ATF collaboration shall submit a notification 
to the Spokesperson, who will relay the matter to ICB.

%% file: cost.tex
\input{cost/cost.estimation.tex}

%% file: cost/cost.estimation.tex
i

In this section 
we discuss the construction expenses 
that are considered necessary to complete the ATF2 
to be ready towards 2008 for starting the initial commissioning. 
The expenses associated with the commissioning operation, 
or possible hardware improvements during commissioning, 
are not accounted for, 
since it is considered premature to evaluate these at this stage of efforts.

The ATF2 will be constructed by an international collaboration 
following the Memoranda of Understanding signed by research institutions participating in the ATF2 project.  

Each research institution will fund all normal operating expenses (salaries, administrative support, and travel ) of its personnel.

The expenses for all of the components are expected to be supported equally by three regions of North America, Europe and Asia where the institutions reside, while the conventional facilities will be funded by KEK as the host institute.  In addition, KEK will fund the operating cost of ATF and ATF2, which is  about $1.5 \times 10^8$yen per year. 

\begin{figure}[htb]
\centering
\includegraphics[width=10cm]{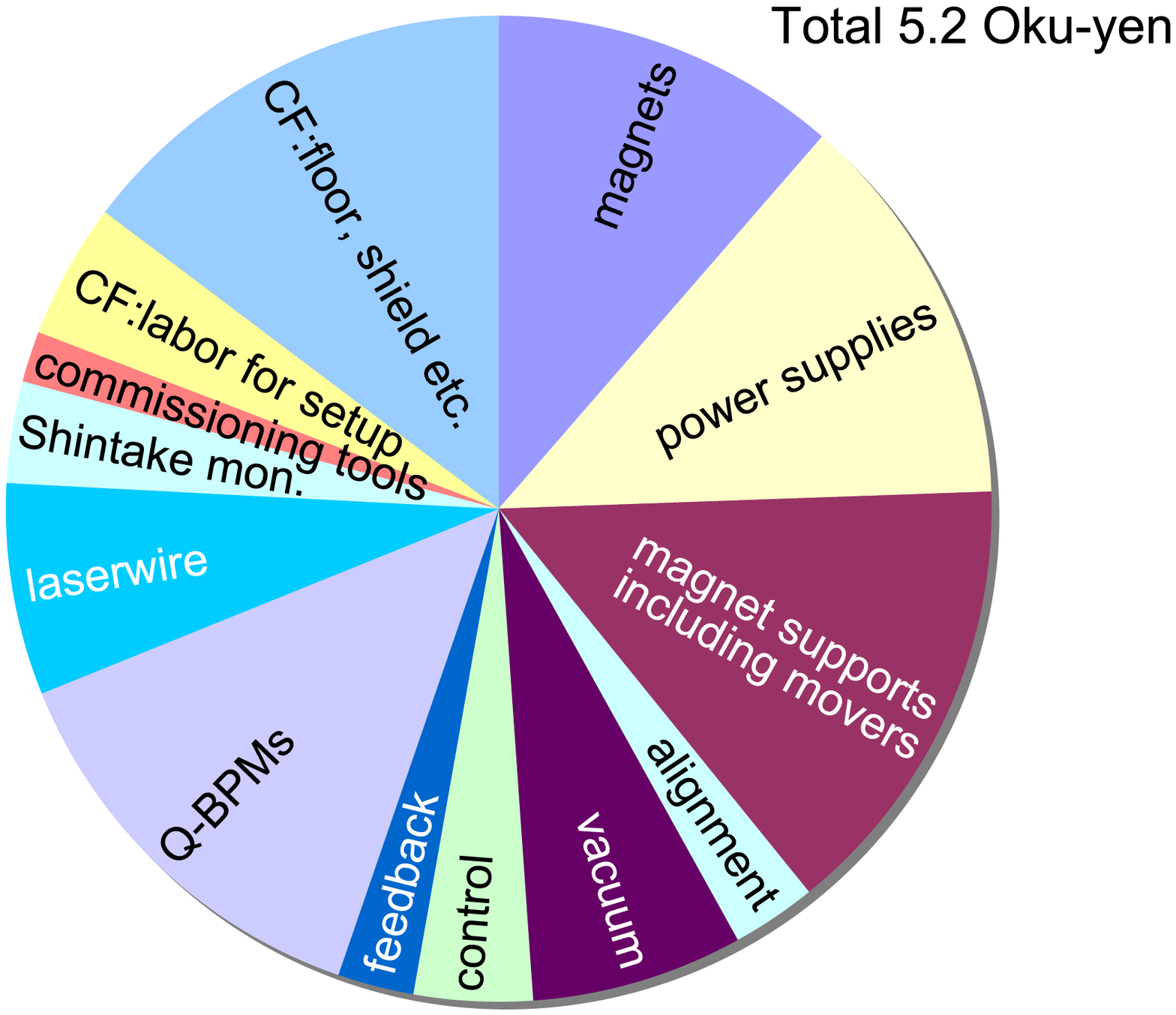}
\caption{Cost distribution of the components normalized by the total cost, where the in-kind ones are also included.}
 \label{cost-cake}
\end{figure}

\begin{figure}[h!]
\vspace{0.5cm}
\centering
\includegraphics[width=13cm]{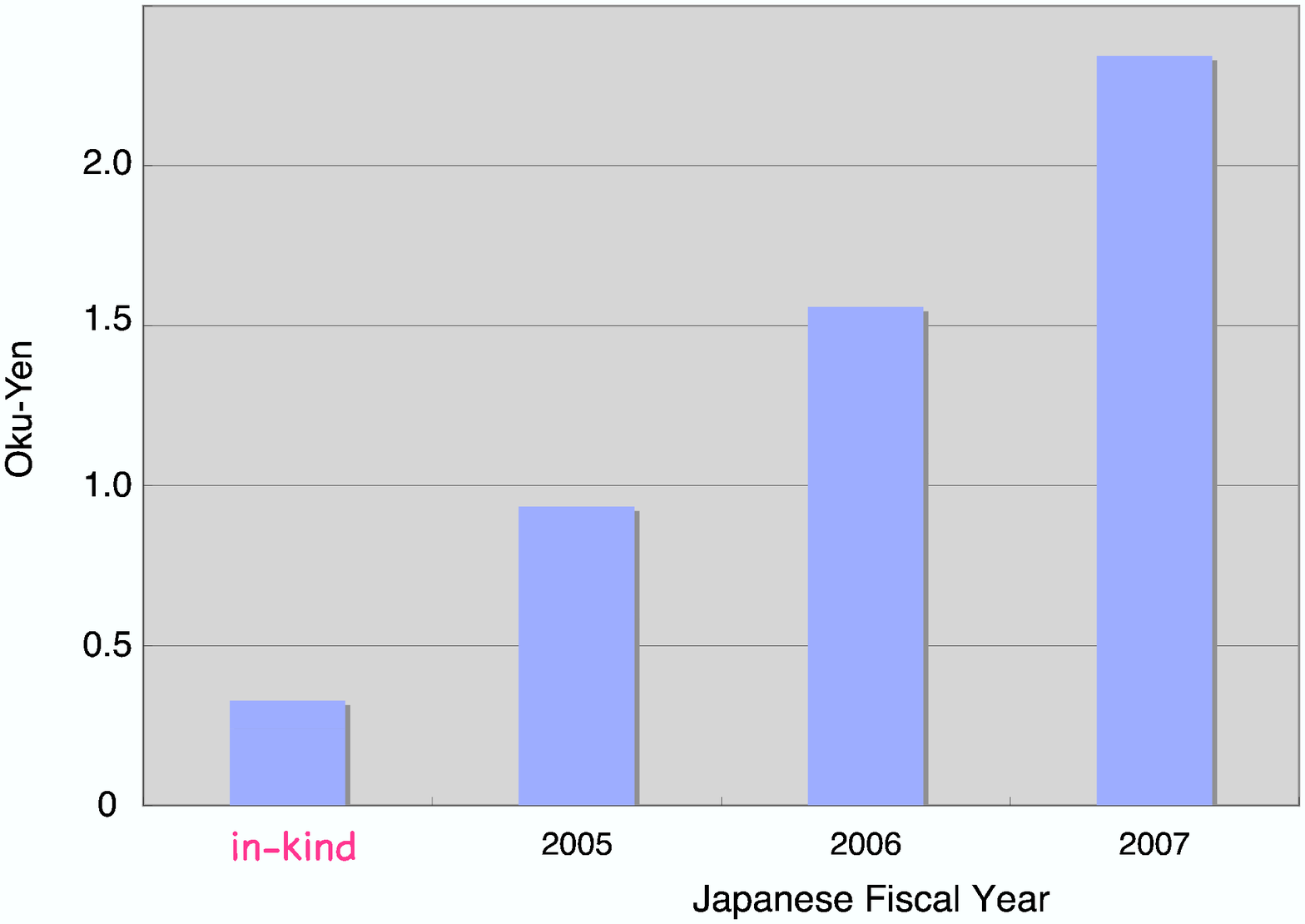}
\caption{Annual budget estimation in  JFY2005 through JFY2007, where in-kind contributions are also shown.}
 \label{annual-budget}
\end{figure}

The cost estimation of the construction is summarized in Table~\ref{tab:cost-estimation}, assuming the maximal use of the ATF facility.
The total cost of the ATF2 project is estimated to be $5.2 \times 10^8$yen.  
No contingency is added, as is usual in a Japanese cost evaluation. 
The cost includes in-kinds, whose estimation is about $0.3 \times 10^8$yen.   
The cost distribution is shown in Fig.~\ref{cost-cake}.   
Annual budgets are estimated to meet the planned commissioning in February, 2008,  
as shown in Fig.~\ref{annual-budget}. 

\begin{table}[htbp]
\caption{Cost estimation for the ATF2 construction.}
\begin{center} 
\begin{tabular}{lc}
&  \\
\hline
Item & Estimated cost \\
        & ($10^8$yen  or Oku yen) \\
\hline
Conventional Facilities& 1.07\\
Magnets& 0.58 \\
Magnet Supports& 0.77 \\
Power Supplies& 0.68 \\
Vacuum& 0.36 \\
Alignment& 0.13 \\
Controls& 0.20 \\
Beam Instrumentation&1.37 \\
\hline 
Total Cost&5.16 \\
\hline 
\end{tabular}
\end{center}
\label{tab:cost-estimation}
\end{table}

%% file: contribution.tex
\input{contribution/tauchi.tex}

%% file: contribution/tauchi.tex

\subsection{Overview}
\input{./contribution/overview.tex}

\subsection{International Contribution}

\subsubsection{IN2P3 in France}
\input{./contribution/orsay.philip.tex}

\clearpage
\subsubsection{DESY}
\input{./contribution/desy.tex}

\subsubsection{CERN}
\input{./contribution/cerncontrib.tex}

\subsubsection{UK LCABD Collaboration}
\input{./contribution/uk.lcabd.tex}

\subsubsection{SLAC}
\input{./contribution/slac.tex}

\subsubsection{IHEP in China}
\input{./contribution/ihep.tex}

\subsubsection{PAL in Korea}
\input{./contribution/pal.tex}

\subsubsection{KEK and University of Tokyo}
\input{./contribution/kek.tokyo-univ.tex}

%% file: contribution/overview.tex
Many institutes have expressed interest of  participating in the ATF2 project, 
as can also be clearly seen in the author list of the ATF2 proposal. 
The contributions of major institutes are summarized in 
Table~\ref{tab:contribution}. 
The listed contributions are preliminary at present, 
since some of them are not yet funded, 
and are still in a process of proposals to their funding agencies.   
The ATF2 project will be completed with these contributions.

The European contribution comes from LAL,LAPP (France), DESY ,CERN and universities in UK.  
They have a strong research program of EUROTeV for the realization of ILC, 
where R\&D on the beam-delivery system is one of major tasks,  as well as the CLIC R\&D program.   
Also, there is the LCABD (Linear Collider Accelerator and Beam Delivery) collaboration in UK.  
Therefore, they will participate in the ATF2 project closely involving them.   
Especially, they expressed the desire to have responsibilities of the laserwire system 
and the feedforward and feedback system for beam-size measurements of the micron-size beam 
and precise orbit control at the sub-micron meter level, respectively.  
CERN has an in-kind contribution of an active support system for the final focus doublet.   
Their active participation in the commissioning and beam operation is also expected.

SLAC is the major laboratory from North America.  
SLAC and KEK have been collaborating on the R\&D of linear colliders 
as well as the B-factory accelerators 
within the framework of Japan-US Science and Technology Cooperative Program.  
The ATF2 project is also one research field in  the program.  
SLAC  will contribute the electronics of cavity BPMs, 
the power supply system as well as the beam tuning and operation.  
Also, SLAC expressed an interest in producing the final doublet system.  
The mover system of magnets is provided in-kind, which have been used at the FFTB, SLAC. 

In Asia, the major laboratories are IHEP (China), PAL (Korea) and KEK as the host. 
IHEP will produce the quadrupole magnets.  
PAL will produce the cavity BPMs.  
They will participate in  the beam tuning and operation as well. 
PAL expressed an interest in the power-supply system, 
since they have recently developed a similar system.   
KEK and the University of Tokyo will take responsibility for the precise  beam-size monitor at IP, 
which is called the Shintake monitor based on the laser interferometer, 
with collaboration with SLAC and UK-universities.  
Also, KEK will contribute the IP-BPM to measure the vertical beam jitter 
with an accuracy of 2~nm at the focus point, 
which will be replaced with the Shintake monitor during the second operational stage.  
As the host, 
KEK will provide all other components,
including the conventional facilities, 
in order to ensure completion within  the proposed schedule.

In Table.~\ref{tab:contribution}, 
there are also human resources estimations 
that can be available at each institute 
during three years of the construction and the commissioning 
for the Japanese fiscal year of 2005 through 2007.  
The human resources are defined by full-time-equivalent (FTE) man times working period,
such as months and years during the three years.   
The total human resources are estimated to be  39.75 person$\cdot$years, 
where the Japanese one is 15 .  In addition to the ATF2 human resources,  
there are 18 person$\cdot$years for ATF operation of the injection and damping ring at KEK.


We will finalize these contributions by April, 2007, 
to meet with the initial commissioning time in February, 2008. 

The following subsections describe the respective contributions in more  detail,
including expressions of interest in  various fields of the ATF2 project.

\input{./contribution/contribution-table.tex}

%% file: contribution/contribution-table.tex
\begin{landscape}
\begin{center}
\tablecaption{International contributions of major institutes,  which are based on the expression of interests.}
\label{tab:contribution}
\tablehead{ 
	\hline \hline
	Item &  
	LAL, LAPP & 
	DESY & 
	CERN & 
	UK & 
	SLAC & 
	IHEP & 
	PAL &
	KEK/U.Tokyo \\
	&  
	(France)& 
	(German) & 
	(Europe) & 
	& 
	(USA) & 
	(China) & 
	(Korea) &
	(Japan)  \\ \hline \hline
}
\tabletail{\hline}
\footnotesize
\begin{supertabular}{p{3.321cm}*{8}{|p{1.853cm}}} \hline%

Mechanical stability and vibrations (include Ground Motion) of Qs & 
	EUROTeV & 
	EUROTeV & 
	1 active stabilization table (2.35$\times$0.8m$^2$ $\times$0.8(H)m) 100kCHF (8.72~Myen) & 
	yes &
	&
	& 
	&
	yes  \\ \hline
GEANT4 modeling of the ATF2 beam line by BDSIM &
	EUROTeV; 1.5~person$\cdot$year at post-doc level in 2006--2008 + travel money &
	&	
	&
	yes &
	&
	&
	&
	\\ \hline  
Evaluation and implementation of steering and optical tuning algorithms, BBA and IP tuning &
	1 PhD student in 2006--2009 + money for visits to KEK &
	&
	7~person$\cdot$month &
	yes+ \ \ \ \ \ \ \ \ \ \ EUROTeV (ILPS) &
	&
	0.5~person$\cdot$year &
	3~person$\cdot$month  &
	yes \\ \hline
Participation in commissioning activities with development of the strategy &
	yes &
	&
	4~person$\cdot$month &
	yes &
	&
	0.5~person$\cdot$year &
	&
	yes \\ \hline
Design of the low-noise electronic readout system, and coherent radiation monitor for beam size measurement &
	technical contributions + money for visits to KEK &
	&
	&
	&
	&
	&
	&
	yes \\ \hline
Fast kicker pulser &
	&
	XFEL &
	&
	&
	&
	&
	&
	yes (ATF) \\ \hline
Laser wire system &
	&
	EUROTeV (LBBD) &
	&
	35~Myen &
	&
	&
	yes
	\\ \hline  
Remote operations &
	&
	EUROTeV &
	&
	&
	&
	&
	&
	\\ \hline
Survey of relevant collective effects in ATF2 and  ATF extraction line
	&
	&
	&
	1~person$\cdot$month (wake field) &
	&
	&
	&
	\\ \hline
Beam feedback / feedforward system &
	&
	&
	&
	4~person/year, 10~Myen &
	4~Myen &
	&
	yes (RF) &
	\\ \hline  
High-precision transformer BPMs (3); 100nm resolution, 4mm aperture
	&
	&
	&
	EUROTeV 160kCHF (13.95~Myen) &
	&
	&
	&
	&
	\\ \hline
QBPM(37):&
	&
	&
	&
	&
	218~k\$ &
	&
	&
	\\
electronics system &
	&
	&
	&
	&
	(24~Myen) &
	&
	&
	\\ \hline
QBPM(37):&
	&
	&
	&
	&
	&
	&
	6~person$\cdot$month, &
	\\
         cavity system &
	&
	&
	&
	&
	&
	&
	41.5~Myen &
	\\ \hline
QBPM(37): &
	&
	&
	&
	&
	&
	&
	&
	5~Myen
	\\
	hybrid + short cables &
	&
	&
	&
	&
	&
	&
	&
	\\ \hline
Q-magnets &
	&
	&
	&
	&
	&
	2~person$\cdot$year, 23.5~Myen &
	&
	\\ \hline
Magnet movers:&
	&
	&
	&
	&
	24~Myen &
	&
	&
	yes\\
	(33 + spares)&
	&
	&
	&
	&
	&
	&
	&
	\\ \hline
Magnet support system (except for FD) &
	&
	&
	&
	&
	&
	&
	&
	44~Myen
	\\ \hline
Power supply system &
	&
	&
	&
	&
	67.5~Myen &
	&
	6~person$\cdot$month (if possible)&
	\\ \hline
Vacuum system &
	&
	&
	&
	&
	&
	&
	&
	38.4~Myen
	\\ \hline
Control system &
	&
	&
	&
	&
	&
	&
	&
	10~Myen
	\\ \hline
Shintake monitor (BSM) &
	&
	&
	&
	&
	3~Myen &
	&
	&
	14.5~Myen  
	\\ \hline
Final doublet &
	&
	&
	&
	&
	5~Myen &
	&
	&
	\\ \hline
Bends: FF and chicane &
	&
	&
	&
	&
	17.5~Myen 
	&
	&
	\\ \hline
6 and 8 pole magnets &
	&
	&
	&
	&
	&
	8.4~Myen &
	&
	\\ \hline
IP-BPM system &
	&
	&
	&
	&
	&
	&
	&
	R\&D \\ \hline
Conventional facility (concrete shields, floor reinforcement)
Utilities (air conditioning, water cooling system, AC power cabling)
Interlock system,
Labor cost (alignment, cabling, setup)&
	&
	&
	&
	&
	&
	&
	&
	107.2~Myen + more \\ \hline
Commissioning tools: &
	&
	&
	&
	&
	&
	&
	yes &
	yes \\
	wire scanners, screen monitors etc. (8.7~Myen) &
	&
	&
	&
	&
	&
	&
	&
	\\ \hline
Alignment: total station system (5~Myen) &
	&
	&
	&
	&
	&
	&
	&
	\\ \hline
FTE\footnotemark for 3 years &
	6.5 &
	? &
	1 &
	24 &
	12 &
	3 &
	1.25 &
	18.3 \\
in JFY2005--2007 &&&&&&&&\\
\hline
FTEs for injection and DR at ATF for 3 years &
	? &
	? &
	? &
	? &
	? &
	? &
	? &
	27 \\
in JFY2005--2007 &&&&&&&&\\
\hline \hline
\end{supertabular}
\footnotetext{FTE = Full Time Equivalent in person$\cdot$year}
\end{center}
\end{landscape}

%% file: contribution/orsay.philip.tex


{\bf Introduction}

At present, 
two IN2P3 laboratories, LAL (contact person Philip Bambade) and LAPP (contact person Yannis Karyotakis), 
are able to contribute to the ATF2 project, as outlined below.

The main technical and scientific motivations have been discussed 
in the context of the ILC-WG4 (Beam Delivery System) activities
in phase with the GDE.
The argument to train young physicists and engineers on a real system is 
also recognized as being important. 

Partial support for French participation to ATF2 should become available 
within the global bilateral Japan-France agreement, 
presently being discussed to promote joint research activities in subatomic physics. 
French participation to ATF2 was among the projects submitted within this framework. 
Although there is no information yet on the level of the potential support, 
the hope is that funds for travel, some extended visits 
and contributions to a few critical short-term contracts 
may become available through this scheme.

The projects that IN2P3 can contribute to towards ATF2 are listed below:

{\bf Mechanical stabilization}

Within the presently funded EuroTeV activities, 
a frequency analysis of the mechanical support structures of the final doublet can be made, 
in view of ensuring appropriate stability. 
This contribution can be connected to the design work of such structures. 
If additional resources become available, 
participation to an active stabilization program can also be considered.

{\bf GEANT4 modeling}

Contributions to a GEANT4 modeling of the ATF2 beam line can be provided, 
based on the BDSIM tool presently used to study the ILC beam delivery system, 
by involving part of the on-going EuroTeV activity at LAL 
(work package ILPS, activity PCDL, 
 which includes GEANT4 modeling of the ILC post-IP extraction line 
 and assessments of the related backgrounds) in 2006. 
This will serve to estimate the backgrounds from beam losses at the dump 
after the "IP" and from a halo produced by upstream collimators. 
In connection with a program of background measurements, 
it will also be critical to validate the BDSIM tool. 
Of particular interest are background conditions 
in the vicinity of the interferometers planned to measure the beam size 
(Shintake, coherent synchrotron radiation). 
Presently, one post-doctoral researcher is being funded throughout 2006 by EuroTeV. 
He can contribute 50\% of his time during 2006. 
To complement this and to enhance the contribution beyond 2006 and into the commissioning period of ATF2, 
a second post-doc has to be recruited by mid-2006 for a 2-year period. He/she should have roughly equal contributions on the ILPC/PCDL ILC extraction line work (as additional LAL contribution to EuroTeV) and ATF2 beam-line GEANT4 modeling. This project is closely connected with the work of the group of Grahame Blair (RHUL,UK), author of BDSIM package and with whom the LAL group has a collaborative agreement through the ALLIANCE bilateral scheme (to fund mutual visits). 
The financing, foreseen for the second post-doc, 
is 50\% from IN2P3 as a direct contribution to the ATF2 project (to be accounted for in 2006-2008 within the funds for the above-mentioned France-Japan collaboration agreement) and 50\% on other laboratory funds (this needs endorsement of the LAL scientific council). 
In total, 
the contribution to this part of the project would thus be 1.5 person-year at the post-doc level during the 2006 -- 2008. In addition, travel money for visits to KEK will be provided through the France-Japan funds. Some complementary travel money for meetings and conferences is available within the LAL ILC group and EuroTeV travel budgets.

{\bf Tuning and commissioning}

Evaluation and implementation of steering and optical tuning algorithms, and participation in commissioning activities. Participation in the development of dedicated beam instrumentation. 
A student has recently started PhD study jointly supervised 
at the University of Valencia (by Angeles Faus-Golfe) and at LAL-Orsay (by Philip Bambade). 
It will last for about 4 years from September 2005,
and a significant part of the thesis work can be spent on ATF2. The funding for this student is secured for the full period (first year at LAL through EU funding, via the research training network contract RTN2-2001-00450, and subsequent years in Valencia with extended periods at LAL, through a newly approved Spanish 3-year contract to fund ILC activities in Valencia and through the bilateral CYCIT-IN2P3 agreement PP05-1, to fund mutual visits). In addition, travel money for visits to KEK will be provided through the France-Japan funds. Some complementary travel money for meetings and conferences is available within the LAL ILC group and EuroTeV travel budgets. For what concerns contributions to beam instrumentation, two projects are considered. The first is to help design the low-noise electronic readout system necessary to reach nanometer-level resolutions with a cavity beam-position monitor at the "IP". The second is to contribute to the implementation of a recently proposed scheme to use coherent synchrotron radiation to monitor the ATF2 beam, based on an idea suggested by Takashi Naito (KEK), adapted from a technique currently used on the ATF ring.

{\bf Summary of contributions to ATF2}

\begin{itemize}
\item Mechanical stabilization (to be specified further by Yannis Karyotakis)
\item GEANT4 modeling of ATF2 beam line: 
    1.5 person-years at post-doc level in 2006-2008 + money for visits to KEK
\item Tuning algorithms, commissioning and instrumentation: 
    1 PhD student in 2006-2009 + money for visits to KEK
\item Instrumentation developments: 
    technical contributions + money for visits to KEK
\end{itemize}

%% file: contribution/desy.tex
{\bf Introduction}

DESY is interested in fostering collaboration with KEK,
specifically in the area of ILC related developments. 
The ATF activities are unique in that they provide small emittance beams
that are required for realistic ILC accelerator and instrumentation tests. 
ATF is planned as an ILC test facility and thus has the benefits of providing flexibility 
for the changing the needs of the experiments as they evolve.
 
At the same time DESY is operating the VUV-FEL as a user facility that grew out of the TESLA test Facility (TTF). 
The prime goal for this facility is to provide FEL beams for the various users; 
however, approximately 1/4 of the time is reserved for accelerator studies, 
which can be viewed as complementary to the ATF program. 
 
The main construction project at DESY during the coming years will be the European X-ray FEL 
based on TESLA SC linac technology.  
Preparation for this ambitious project currently dominates DESY's R\&D program, 
and through the synergy of the shared linac technology represents DESY's largest contribution to the ILC.  
Given this situation, it is clear that DESY will not have large resources 
available to invest in additional or unrelated activities. 
 
The ILC group at DESY tries to benefit from the activities for the VUV-FEL and the XFEL 
in an optimal manner and contributes to the XFEL in some fundamental research. 
There are some additional resources that DESY sets aside for ILC specific research. 
In addition, DESY contributes to the ILC effort via its participation in the EUROTeV program.
Given the current commitments and limited resources, it is only possible for DESY 
to collaborate on new projects that fit well within existing programs and goals. 
It will not be possible for DESY to assume new larger responsibilities at this time. 
 
{\bf Participation in Experiments at ATF}
 
DESY wishes to take part in the experimental program at the ATF. 
DESY wishes to emphasize the parts of the program that are complementary to the international program at the TTF. 
In particular DESY wishes to
\begin{itemize}
\item contribute to the running of ATF (shifts, preparation, data analysis)
\item profit from the experience of using new instrumentation and the possibility to test and use new equipment
\item help to propose experiments at the ATF in a worldwide coordinated fashion 
while taking the available resources into account
\end{itemize}

{\bf Fast Kicker Pulser}

DESY is interested to advance the pulser development for the fast kickers 
in collaboration with KEK. 
Such a program is already under way (F. Obier), and is mandatory for the XFEL. 
It is possible that the same pulser can be used for the XFEL and for ATF. 
As part of the existing collaboration, 
DESY wishes to continue measurements of the rise and fall times of the kicker pulse using the ATF beam. 
After successful tests and studies DESY may be in a position 
to construct and supply the required kicker pulser modules for ATF, 
providing the necessary resources can be made available.

{\bf Laser Wire Facility}

DESY is part of the Laser Based Beam Diagnostics (LBBD) collaboration, 
through its support of the laserwire facility at PETRA. 
DESY's contribution to this project has now increased 
with the purchase of a new laser (within the EUROTeV program). 
DESY would be interested to extend its current program 
to the new facilities planned for ATF2 in a manner 
that is compatible with the the EUROTeV program, together with our UK collaborators.

{\bf Remote Operations}

DESY is participating in the development of a Mobile Virtual Laboratory as part of its EUROTeV program. 
The activities are coordinated by F.Willeke and will serve to enable remote diagnostics and interferences. 
The ATF is planned as an international user facility, 
and could thus serve as a showcase for the functionality of the approach.
DESY is interested to extend these activities towards a remote facility, such as the ATF. 
Other institutions would be welcome to engage in the remote operation.

{\bf Mechanical Stability and Vibrations}

DESY has been taking ground motion measurements at many sites worldwide. 
DESY is also engaged in characterizing the ground motion and so-called `cultural noise'. 
A program to develop accurate vibration sensors 
to measure spectra in the cold mass of the linac cryomodule is also underway. 
The work is carried out as part of DESY's EUROTeV studies and is complemented by activities in Oxford and Annecy. 
Such R\&D could be extended to the ATF2 program.

In addition it will be necessary to characterize the need for active stabilization at the ILC. 
The ATF would be an ideal place to counterbalance the need for sophisticated stabilization 
at the sub-micron level versus the benefits of active feedback systems and control loops as they are foreseen for the ATF.

{\bf Summary}

DESY is seeking to increase the collaboration 
with KEK within the constraints of its limited free resources. 
DESY wishes to take part in the experimental program at the ATF and to contribute to its analysis. 
DESY has made a proposal for specific participation 
in the areas of kickers, laser wires, remote control, ground motion and stabilization.

%% file: contribution/cerncontrib.tex


{\bf Introduction}

Since additional resources have been requested for the
CLIC Test Facility No.~3 (CTF-3), 
the CLIC Team can presently offer as CERN contributions to the ATF2 Project  
only existing equipment and studies. 

{\bf Studies} 

The CLIC Team proposes 
contributions to the following four groups of studies:
\begin{itemize}
\item 
Development of a commissioning strategy.
A corresponding section was already
written for the ATF2 design report, Volume 1 \cite{atf2vol1}.
\item 
Investigation of optimum beam-based alignment
procedures, e.g., as in \cite{alignmentdaniel}, 
with pertinent specification of 
the BPM ranges.
\item 
Simulations of IP tuning and the determination of the maximum tuning knob ranges, 
e.g., in the spirit of \cite{nima}. 
\item 
Survey of relevant collective effects in ATF2
and the ATF extraction line, in particular wake fields,
e.g., as in \cite{slcwf}. 
\end{itemize}
 
{\bf Hardware} 

In addition, the following hardware contributions are proposed.
\begin{itemize}
\item
An active stabilization table including stabilizing feet
(STACIS2000 system from TMC; value about 100 kCHF). 
The table is a honeycomb support structure with length 2.4 m,
width 0.8 m, and height 0.8 m. The manufacturer guaranteed 
the absence of any structural resonances below 230 Hz. 
The stabilizing feet are equipped with 
integrated geophones for measuring ground
vibrations, rubber pads for passive damping, and piezoelectric
movers for active damping of load vibrations induced from
the ground. The stabilized table was successfully used for the 
CERN stability study \cite{cernstability}. Presently it 
is on loan at LAPP (Annecy). The table would be available
from the middle of 2006. The best use of this table
for ATF2 needs to be identified. One possibility
may be to stabilize the final quadrupoles and the 
IP monitors. 
\item
Three high-precision transformer BPMs 
(development costs about 160 kCHF plus
human resources, co-financed by the European Union
within EUROTeV) \cite{bpm}. The projected spatial BPM
resolution is 100~nm and the time resolution 15~ns. 
The nominal BPM aperture is 4~mm, 
but it could be increased to, e.g., 6~mm.
These BPMs would be available at the end of 2007. 
The starting point of the EUROTeV BPM design 
is a similar monitor, used at CTF-3, with a 
40-mm aperture \cite{gasior}. The goal of 
the EUROTeV work package \cite{soby} is to reduce
the size of the BPM by a factor 
of 10 and to improve its resolution
accordingly, from a few microns to 100~nm.
The advantages of the transformer BPMs 
compared with, e.g., rf BPMs, are 
that they function for arbitrary 
bunch spacing, have a large frequency
bandwidth (almost 6 decades) and a simple readout
electronics, and that they are insensitive to 
spray from lost beam particles.
The EUROTeV packages leaves some flexibility.
For the ATF/ATF2 application, information 
is still needed on the desirable aperture, the
surrounding beam pipe, the flange sizes,
and the type of flanges.

\end{itemize}









%% file: contribution/uk.lcabd.tex

The UK LCABD (Linear Collider Accelerator and Beam Delivery) Collaboration comprises 14 UK institutes working mainly in aspects
of the ILC Beam Delivery System. A number of institutes are already
involved in collaborative R\&D at ATF, and are interested in participating
in ATF2. A brief outline of the areas of interest is:
\begin{itemize}
\item 
 Development of steering and optical tuning knobs and algorithms
(Daresbury Lab.)
\item 
 Simulations of IP tuning and optimization of tuning knobs
(Daresbury, QMUL)
\item 
  Commissioning strategy and participation in commissioning activities
(Cambridge, Daresbury Lab., Oxford, QMUL, RHUL, UCL)
\item 
   Development of a laserwire system for ATF(2)
(Oxford, RHUL, UCL)
\item 
    Development of a beam-based intra-train feedback system for beam
stabilization for ATF(2)
(Daresbury Lab., Oxford, QMUL)
\item 
     Development of a ring-to-linac feed-forward system for stabilization of
the vertical beam position in the ATF(2) extraction line
(Daresbury Lab., QMUL)
\item 
      Development of optical alignment capability for ATF NanoBPMs, and
final-focus magnets in ATF2
(Oxford)
\item 
       Participation in NanoBPM development for ATF(2)
(Cambridge, UCL)
\end{itemize}

The current LCABD project is funded until April 1st, 2007. 
Contributions can be made to ATF(2) within the framework of the approved LCABD program.
UK funding for any contributions beyond April 1st, 2007 could be applied for
when the UK funding agencies (PPARC, CCLRC) make their call for proposals,
which we currently expect to be in mid-2006. Some UK BDS activities are
also supported within the EUROTeV framework.

%% file: contribution/slac.tex
KEK and SLAC have a long history of fruitful collaboration, 
and have worked jointly on ATF since the design phase.  
The ATF2 project will bring this collaboration to a new international level. 
SLAC is interested in participating in design studies and in developing and contributing to ATF2 hardware 
similar to that needed for ILC.  
SLAC is also interested in contributing hardware existing at SLAC for reuse at ATF2 and 
participating in ATF2 commissioning and operation.

Design studies for ATF2:
\begin{itemize}
\item Design studies for ATF2 optics
\item Evaluation of wake-field effects
\item Analysis of tolerances
\item Simulation of tuning methods
\item Simulation of alignment system
\item Design of extraction from ATF ring
\end{itemize}

Development of ILC-like hardware and other hardware for ATF2:
\begin{itemize}
\item Electronics for 100~nm resolution ATF2 beamline BPMs 
\item ATF ring BPMs with better resolution and lower systematics 
\item High availability, stable power supplies for ATF2 beamline magnets
\item High availability fast kicker pulsers
\item Final doublet quadrupole magnets
\item Dipole bending magnets
\item Participate in the design and production of the ATF2 beamline quads at IHEP
\item Precision magnets movers for ATF2 
\end{itemize}
Contribution of personnel power for commissioning and operation:

\begin{itemize}
\item Participate in operation and commissioning of ATF2
\end{itemize}

An estimate of the personnel-power contribution from SLAC to ATF2 is expected to be approximately 12 FTEs  in JFY 2005-2007 (3 years);
however the specific hardware development and contributions from SLAC to ATF2 depend 
details of the collaboration agreements and funding levels.

%% file: contribution/ihep.tex
 
 
The Institute of High Energy Physics (IHEP), Chinese Academy of Sciences, is interested 
in strengthening the collaboration with KEK within ILC-Asia and ILC-Global frames.
ATF2 
as one of the most important international collaboration R\&D projects for ILC 
provides a unique possibility 
to study very low-emittance electron beam, 
with its production, manipulations, instrumentation, measurements, and operation. 
IHEP is working on an R\&D program of XFEL. 
There are many common physical and technical problems for ILC (ATF2) and XFEL. IHEP takes the participation to ATF2 as one opportunity to accumulate the experiences for advanced accelerator technology for future relevant projects.
 
 
IHEP is fabricating 24 quadrupoles and may fabricate other magnets for the ATF2.
The specification is listed in the Table \ref{tab-qmagnet}. The field measurements will be conducted at IHEP.
 
\begin{table}[htb]
\caption{Specification of the quadrupole magnet.}
\begin{center}
\begin{tabular}{ll}
\hline
\hline
\bf{Quadrupole Type}& \bf{QEA-D32T180}  \\
 Maximum integrated strength (Tesla) & 10.879 \\
 Predicted effective length (meters) & 0.1978 \\
 Maximum gradient (Tesla/meter) [based on predicted L eff] & 55.0\\ \hline
\multicolumn{2}{l}{{\bf Main Coils} (Values are per magnet)} \\
 Maximum Current (Amps)  &130.0 \\
 Maximum Power (kW)   &1.98 \\
  Resistance at 40$^o$C ($\Omega$)  &0.117 \\
  Incoming cooling water temperature ( $^o$C)  &25.4$\pm$ 0.1 \\
  Inlet water pressure (Kg/cm$^2$)  & 7.14\\
  Design Water p (Kg/cm$^2$) across TWO water circuits  &6.0 \\
 Total Water Flow at Design p (litres/min) at $\sim$ 20$^o$C   &2.2 \\ \hline
\multicolumn{2}{l}{{\bf Trim Coils} (Values are per magnet)} \\
  Maximum Current ( DC Amps)  & 5.0\\
  Maximum Power (W)  &6.2 \\
  Resistance at 40$^o$C (m$\Omega$)  &247\\
  \hline
  \hline
  \end{tabular} 
\end{center}
\label{tab-qmagnet}
\end{table}

From the planning, the quadrupole magnet fabrication will start for ATF2, right after the magnet fabrication work for BEPC-II is finished.

Concerning the beam dynamics in the ATF2 beam line, 
IHEP accelerator physicists wish to participate 
in the beam line design, installation, tuning algorithms, commissioning, data taking and analyzing. 
IHEP can contribute to the running of the ATF, 
such as shifts, machine studies and data analysis. 
Advanced instrumentation technologies are subjects of interests for IHEP, 
and the involvement of relevant task forces is foreseen in the future. 

%% file: contribution/pal.tex

ATF2 is a promising test facility to study the beam generation, beam dynamics, beam diagnostics, 
and beam operation for a low-emittance and short-bunched Linac, such as the ILC. 
Most of the ILC technology will also be applicable to the X-ray FEL. 
As the PAL (Pohang Accelerator Laboratory) is pursuing 
both the XFEL program and the ILC collaboration, 
the participation of scientists in the construction and beam study of ATF2 should be beneficial to PAL 
by boosting the synergy of the shared technology. 
Collaborations with other worldwide participating laboratories are also desirable. 
Possible contributions to ATF2 will be beam-dynamics simulations, 
beam instrumentation (Cavity BPM), 
rf system (phase measurement and feedback) and precision magnet power supply technology. 
Total of 15 person-month human-resource contributions are foreseen. 
The traveling expenses for visits to KEK can be benefited 
by the utilization of partial support from Japan-Korea CUP (Core University Program).

\clearpage
{\bf Beam Physics}

PAL is a current member of the ILC design effort group. 
Bunch compression and the damping ring design for ILC are major activities for now. 
Participation in beam parameter measurements, feedbacks 
and beam-based optimization of ATF2 should be considered. 
The contribution of 3 person-month human resources for this study are foreseeable.  

{\bf Beam Instrumentation}

As an active part of the PAL contribution to the ATF2 program, 
PAL is fabricating cavity BPMs for ATF2. 
The required detection resolution of the cavity BPM for ATF2 is better than 100~nm. 
With expertise in the precision machining and fabrication in PAL, 
we have successfully completed fabrication and cold rf tests of a prototype cavity BPM. 
Cold tests of the BPM characteristics, 
such as the resonance frequency, Q-factors and isolation between vertical and horizontal sensitivities, 
can be performed in PAL using a network analyzer. 
We are also interested in a hot test of BPMs with a real beam in ATF2 beamlines. 
The contributions of 6 person-months of human resources will be needed for cavity BPM collaboration. 
Other kinds of beam-diagnostic monitors, 
such as a wire scanner and laser wire, 
are also subjects of interest to us. 

{\bf Magnet Power Supply}

Technology for a precision magnet power supply is another unique area 
that PAL can contribute to an advanced accelerator program, like ATF2. 
Depending on the hardware work, 
up-to 6 person-months of human resources can be allocated for the job. 
Activities that have been done in PAL are: 
\begin{itemize}
\item development and operation of $\pm$110A digital bipolar power supplies
  with accuracies of 2~ppm (10~min) and 10~ppm (12~hr)
\item development and operation of $\pm$30A analog switching power supplies 
  with accuracies of 5~ppm (10~min) and 50~ppm (12~hr)
\end{itemize}
  
{\bf RF Technology}

A pulse-to-pulse phase measurement and feedback technique for the rf system is also foreseen 
as a possible contribution to ATF2 operation. 
A digital phase measurement and analog phase shifter comprises an rf feedback system. 
The phase accuracy of the measurement achieved in PAL is 0.03 degree-rms.

%% file: contribution/kek.tokyo-univ.tex
 
The total operational time of ATF was 21 weeks  with 110 hours/week  in JFY2004.   Twenty five foreign researchers  have conducted experiments at ATF, which took about 30\% of the beam time.    ATF has been annually operated with 9 FTEs (Full Time Equivalent) comprising of 3, 3 and 3 FTEs for the injection, damping ring and extraction line, respectively.   The operation was fully supported by KEK. 
 
Since ATF2 is an extension of ATF to be fully integrated, 
the above-mentioned human resources will also be available at ATF2. 
In addition, 5 FTEs will contribute to following activities each year in JFY 2005 -- 2007: 
i.e. the construction period. 
 
KEK contributes in the optics design and an estimation of the tolerances for the magnet stabilities, 
such as the field strength and the positions, 
while the originally proposed optics has been evaluated 
by comparing to the adopted ILC scaled optics.  
Planning of the commissioning strategy and the tuning methods  is also an active field, 
involving many years of experience in the ATF operation at KEK. 
 
KEK closely collaborates with IHEP and SLAC 
to design quadrupole magnets and to evaluate the performance 
by magnetic field measurements.  
 
The magnets must be stabilized, 
especially in the vertical direction at sub-micron meter level for the final focus beam line, 
whose requirements are similar to the ILC.  
KEK produces the magnet support system using FFTB movers.  
The system can be applicable at ILC.
 
Cavity BPM (Q-BPM) is one of essential instruments 
for precise beam-position measurements 
with an accuracy of 100~nm at the ILC. 
In close collaboration with PAL,
KEK participates in the design of Q-BPM 
and in the performance measurement at the ATF beam line
as well as at a test bench.
Also, KEK develops and produces  the IP-BPM system with an accuracy of 2~nm, 
which measures the vertical jitters of the focal point.
 
All of the ATF2 components are incorporated in the upgraded ATF control system.   
The vacuum system is also provided by KEK.
 
As already mentioned, 
KEK takes care of the conventional facility, 
including the reinforcement of floors, concrete shields and the utilities, 
such as the water-cooling system, air-conditioning and interlock system. 
 
The University of Tokyo has the prime responsibility for a precise beam-size monitor, 
called a Shintake monitor, in close collaboration with KEK, SLAC and UK-universities.   
The Shintake monitor has been used at the FFTB/SLAC, and it will be upgraded.   
The major upgrades are the optical system, 
including the laser, itself, 
with a half wavelength of 532~nm and precise phase detection and control, 
as well  as the detector system for measuring scattered photons.  
Background simulations are also active research items
in collaboration with the optics group for optimizing the collimation system in the beam line.

%% file: postscript.tex
\sect{Concluding remarks}\label{sec:concluding}
\input{postscript/concluding.tex}


\cleardoublepage
\input{postscript/reference.tex}

%% file: postscript/concluding.tex

The ATF2 is an indispensable component of the International R\&D program 
towards the ILC, the International Linear Collider. 
As described in this document, 
it represents a scaled model of the Beam Delivery System, BDS,  
which transports, focuses and controls the beams into the collision points 
within the detectors. 
In the ATF2 presented here, 
the collaboration has produced a system 
designed to test the optics, tolerances, instrumentation 
and controls to demonstrate a proof of principle for the ILC.
The ATF2 is unique in that beginning with the ultra-low emittance beam from the ATF, 
its beam stability, size and instrumentation will mimic that of the ILC but obviously at lower energy.

The Coordination Group formed by the collaboration, 
and described above, has proved to be effective 
in producing the technical design, the construction and installation plans, 
the associated budgets, schedules and defined responsibilities of the partners. 
This team,
with the continuing support of the international community, 
will demonstrate the BDS design, 
offer learning opportunities on the construction and operation of BDS to all who are interested, 
particularly the younger members within the community, 
and be a critical contributor to the realization of the ILC.


\hspace{3.7in}Ewan Paterson
\vspace{-0.2cm}

\hspace{3.7in}Chair International Collaboration Board

%% file: postscript/reference.tex
\newpage